\documentclass[aip,prl,reprint,superscriptaddress]{revtex4-2} 

\usepackage[T1]{fontenc}
\usepackage{times}

\usepackage{graphicx}
\usepackage{booktabs}

\usepackage{amsfonts, amsmath,amssymb, bm, graphicx}
\usepackage{siunitx}
\DeclareSIUnit{\belmilliwatt}{Bm}
\DeclareSIUnit{\dBm}{\deci\belmilliwatt}

\usepackage[english]{babel}

\newcommand{\figref}[2]{\hyperref[#1]{\ref{#1}(#2)}}
\newcommand{\figrefsub}[3]{\hyperref[#1]{\ref{#1}(#2)#3}}

\usepackage{physics}
\usepackage{lipsum}
\usepackage{changes}

\usepackage[colorlinks=true,allcolors=blue]{hyperref}
\usepackage{footmisc}

\usepackage{upgreek}

\usepackage{soul}

\makeatletter
\let\ORIbbl@fixname\bbl@fixname
\def\bbl@fixname#1{%
  \@ifundefined{languagealias@\expandafter\string#1}
    {\ORIbbl@fixname#1}
    {\edef\languagename{\@nameuse{languagealias@#1}}}%
}
\newcommand{\definelanguagealias}[2]{%
  \@namedef{languagealias@#1}{#2}%
}
\makeatother

\definelanguagealias{en}{english}

\begin{document}

\title{Self-induced Floquet magnons in magnetic vortices}

\author{C. Heins}
\affiliation{Helmholtz-Zentrum Dresden--Rossendorf, Institut f\"ur Ionenstrahlphysik und Materialforschung, D-01328 Dresden, Germany}
\affiliation{Fakult\"at Physik, Technische Universit\"at Dresden, D-01062 Dresden, Germany}

\author{L. K\"orber}
\affiliation{Helmholtz-Zentrum Dresden--Rossendorf, Institut f\"ur Ionenstrahlphysik und Materialforschung, D-01328 Dresden, Germany}
\affiliation{Fakult\"at Physik, Technische Universit\"at Dresden, D-01062 Dresden, Germany}
\affiliation{Radboud University, Institute of Molecules and Materials, Heyendaalseweg 135, 6525 AJ Nijmegen, The Netherlands}

\author{J.-V. Kim}
\affiliation{Centre de Nanosciences et de Nanotechnologies, CNRS, Universit\'e Paris-Saclay, 91120 Palaiseau, France}

\author{T. Devolder}
\affiliation{Centre de Nanosciences et de Nanotechnologies, CNRS, Universit\'e Paris-Saclay, 91120 Palaiseau, France}

\author{J. H. Mentink}
\affiliation{Radboud University, Institute of Molecules and Materials, Heyendaalseweg 135, 6525 AJ Nijmegen, The Netherlands}

\author{A. K\'akay}
\affiliation{Helmholtz-Zentrum Dresden--Rossendorf, Institut f\"ur Ionenstrahlphysik und Materialforschung, D-01328 Dresden, Germany}

\author{J. Fassbender}
\affiliation{Helmholtz-Zentrum Dresden--Rossendorf, Institut f\"ur Ionenstrahlphysik und Materialforschung, D-01328 Dresden, Germany}
\affiliation{Fakult\"at Physik, Technische Universit\"at Dresden, D-01062 Dresden, Germany}

\author{K. Schultheiss}\email{k.schultheiss@hzdr.de}
\affiliation{Helmholtz-Zentrum Dresden--Rossendorf, Institut f\"ur Ionenstrahlphysik und Materialforschung, D-01328 Dresden, Germany}

\author{H. Schultheiss}\email{h.schultheiss@hzdr.de}
\affiliation{Helmholtz-Zentrum Dresden--Rossendorf, Institut f\"ur Ionenstrahlphysik und Materialforschung, D-01328 Dresden, Germany}

\date{\today}


\begin{abstract}
Driving condensed matter systems with periodic electromagnetic fields can result in exotic states not found in equilibrium. Termed Floquet engineering, such periodic driving applied to electronic systems can tailor quantum effects to induce topological band structures and control spin interactions. However, Floquet engineering of magnon band structures in magnetic systems has proven challenging so far. Here, we present a class of Floquet states in a magnetic vortex that arise from nonlinear interactions between the vortex core and microwave magnons. Floquet bands emerge through the periodic oscillation of the core, which can be initiated by either driving the core directly or pumping azimuthal magnon modes. For the latter, the azimuthal modes induce core gyration through nonlinear interactions, which in turn renormalizes the magnon band structure. This represents a self-induced mechanism for Floquet band engineering and offers new avenues to study and control nonlinear magnon dynamics. 
\end{abstract}

\maketitle

The electronic band structure of a crystal is characterized by Bloch states, which reflect the discrete translational symmetry of the underlying periodic potential in space. For periodic driving in time, an analogous phenomenon called Floquet states can arise. While Bloch states are shifted in momentum space, Floquet states are shifted in energy by multiples of the drive frequency, which expands the range of possible behavior and properties of condensed matter~\cite{oka_floquet_2019}. Recently, periodic drive using ultrafast laser pulses has been used to induce topological Floquet states~\cite{oka_photovoltaic_2009, lindner_floquet_2011, wang_observation_2013, rechtsman_photonic_2013, mahmood_selective_2016, hubener_creating_2017, rudner_band_2020, mciver_light-induced_2020}, Floquet phase transitions \cite{weidemann_topological_2022, hubener_phonon_2018}, modulations of optical nonlinearity \cite{shan_giant_2021}, novel states in Josephson junctions \cite{park_steady_2022}, and to perform band engineering in black phosphorus~\cite{zhou_pseudospin-selective_2023}. Similarly, Floquet states also enable the dynamical control of spin exchange interactions~\cite{mentink_ultrafast_2015, mikhaylovskiy_ultrafast_2015, itin_effective_2015, claassen_dynamical_2017, kitamura_probing_2017, liu_floquet_2018}, which suggests the possibility of inducing novel features in the collective excitations of magnetically ordered systems, such as magnons.

While Floquet engineering in magnetic systems has been studied theoretically in different contexts~\cite{sato_laser-driven_2016, owerre_floquet_2017, owerre_floquet_2018, owerre_magnonic_2019, elyasi_topologically_2019, nakata_laser_2019, bostrom_light-induced_2020, yang_theory_2023, shi_interacting_2023}, experimental evidence of magnetic Floquet modes remains scarce. The main difficulty in using laser illumination, for example, to modulate intrinsic material parameters such as exchange and anisotropy is that strong dissipation in the electron and phonon systems occurs much faster than the characteristic time scale of coherent excitations, such as magnons in the microwave regime. Here, we present an approach for magnetic materials that does not involve modulating material constants directly, but instead harnesses distinct internal modes that act as the periodic drive. Specifically, we show that the sub-GHz gyrotropic eigenmode of a magnetic vortex~\cite{shinjo_magnetic_2000}, as sketched in Fig.~\ref{fig:geometry}, can induce Floquet states through nonlinear coupling. These states are distinct from the GHz-range magnon modes about a static vortex at equilibrium. Moreover, the sole excitation of GHz magnons by an external magnetic field drives the vortex core into steady state gyration through this nonlinear coupling, which in turn renormalizes the magnon band structure. We term this \emph{self-induced} Floquet band engineering.

\section*{Ground-state magnons of a magnetic vortex state}

    \begin{figure}
    \includegraphics{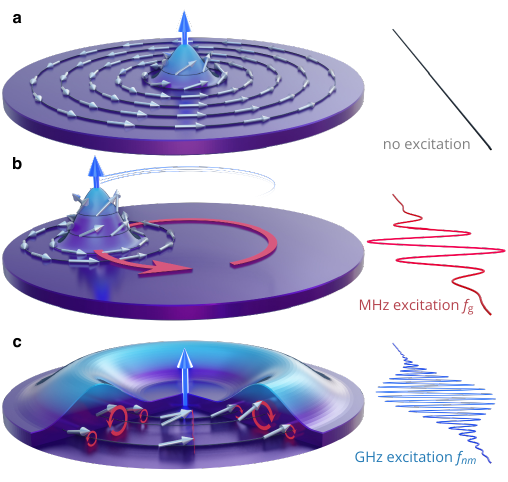}
    \caption{\textbf{Excitations of the magnetic vortex state}. \textbf{a}, Vortex ground state in a magnetic disk, a few tens of nm in thickness and a few $\mu$m in diameter, with the magnetic moments (white arrows) curling in-plane around the vortex core (blue arrow), a narrow region at the disk centre where the magnetization points perpendicularly to the film plane. \textbf{b}, Gyration of the vortex core around its equilibrium position, driven by oscillatory external fields in the MHz range, $f_\mathrm{g}$. \textbf{c}, Regular magnon modes with frequency $f_{nm}$ in the GHz range, where the vortex core remains quasi-static in the disk centre. Here, the lowest-order mode with radial index $n=0$ and azimuthal index $m=0$ is sketched. Displacements and amplitudes are not drawn to scale.}
    \label{fig:geometry}
    \end{figure}

Ferromagnetic disks with certain aspect ratios host magnetic vortices as ground states~\cite{scholz_transition_2003}, as shown in Fig.~\figref{fig:geometry}{a}. Vortices possess two distinct classes of dynamical eigenmodes, gyrotropic and geometrically-quantized magnon modes. Gyrotropic modes involve the gyration of the vortex core around its equilibrium position at the disk center~\cite{novosad_magnetic_2005, stoll_imaging_2015} [Fig.~\figref{fig:geometry}{b}]. The frequency of the fundamental gyration mode is in the MHz range and proportional to the geometric aspect ratio $f_\mathrm{g} \sim L/D$ to lowest order, where $D$ and $L$ are the diameter and thickness of the disk, respectively~\cite{guslienko_eigenfrequencies_2002}. For quantized magnon excitations, the core remains quasi-static in the center and the magnetic moments in the skirt of the vortex precess collectively~\cite{guslienko_vortex-state_2005, vogt_optical_2011}. In thin disks, these quantized modes are in the GHz range and are indexed by two integers ($n,m$), where the radial index $n$ denotes the number of nodal lines along the radial direction of the disk, while the azimuthal index $m$ counts the number of periods along the angular direction. Figure~\figref{fig:geometry}{c} depicts the fundamental mode ($0, 0$). The frequencies of both classes of modes strongly depend on the material parameters and disk dimensions.

\begin{figure*}
    \includegraphics{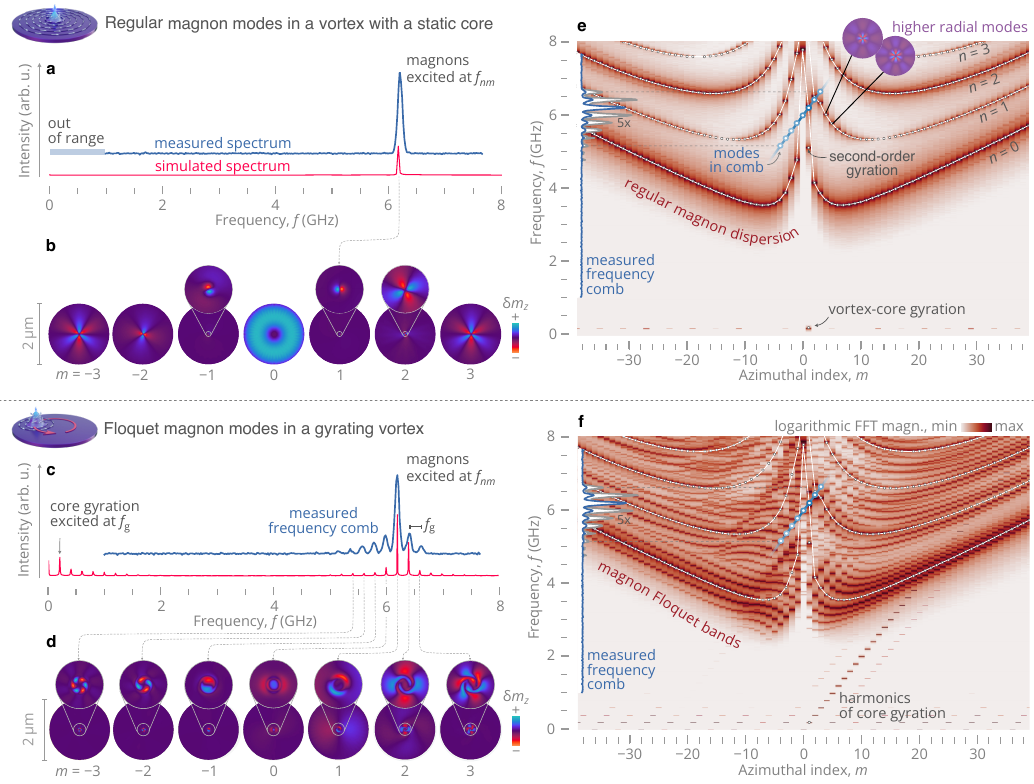}
    \caption{\textbf{Regular magnon modes in a vortex with a static core compared to Floquet magnon modes in a gyrating vortex.} \textbf{a}, Spectra of the magnetic response in a \SI{50}{\nano\meter}-thick and \SI{2}{\micro\meter}-wide Ni$_{81}$Fe$_{19}$ disk when exciting at a single microwave frequency of $f_{nm}=\SI{6.2}{\giga\hertz}$ and the vortex core remaining static. The upper blue spectrum is measured using BLS microscopy (at \SI{-5}{dBm}), the lower red spectrum results from micromagnetic simulations (at  \SI{0.25}{\milli\tesla}). Curves are shifted vertically for clarity. \textbf{b}, The simulated spatial profile of the dynamic out-of-plane component $\delta m_z$ of the mode excited at $f_{nm}=\SI{6.2}{\giga\hertz}$ confirms that only the mode ($0,1$) responds to the drive. The other modes with $n=0$ and different $m=0, -1, \pm2, \pm3$ are not part of the spectral response but only plotted for illustrating their spatial profiles.  \textbf{c}, Simultaneously driving the vortex core gyration at $f_\mathrm{g}=\SI{200}{\mega\hertz}$ and the magnon mode at $f_{nm}=\SI{6.2}{\giga\hertz}$ leads to the generation of a magnon frequency comb with its spacing matching $f_\mathrm{g}$. \textbf{d}, The simulated profiles of the comb modes in this gyrating vortex show clear differences from their regular counterparts in the vortex with a static core. \textbf{e}, Simulated dispersion of thermally populated magnon modes in the vortex with a static core plotted for different radial indices. The modes measured in the frequency comb (hollow blue dots) do not coincide with modes on the regular magnon dispersion. \textbf{f}, Micromagnetic simulations of the thermal magnon population in a gyrating vortex show the Floquet magnon bands. To highlight their differences, the white data represents the regular magnon dispersion.}
    \label{fig:modes}
    \end{figure*}

We studied vortex modes in ferromagnetic Ni$_{81}$Fe$_{19}$ disks patterned on top of a \SI{2}{\micro\meter}-wide central signal line of an on-chip coplanar waveguide. Microwave currents flowing through this waveguide generate oscillating in-plane magnetic fields, which due to their symmetry couple directly only to either the vortex gyration or to magnon modes with azimuthal mode numbers $m=\pm 1$, depending on the applied frequency, as shown in Fig.~\figref{fig:modes}{a}. For a \SI{2}{\micro\meter}-diameter disk, a strong resonant response can be observed for a microwave field at $f_{nm}=\SI{6.2}{\giga\hertz}$, which is visible in the experimental spectra obtained with Brillouin light scattering (BLS) microscopy (for a microwave power of \SI{-5}{dBm}) and micromagnetic simulations (for an excitation amplitude of \SI{0.25}{\milli\tesla}). Fig.~\figref{fig:modes}{b} shows the simulated spatial profile of the mode excited at $f_{nm}=\SI{6.2}{\giga\hertz}$, which confirms that the azimuthal mode ($0, 1$) couples effectively to the microwave drive. Simulated profiles for modes with $n=0$ and different $m=0, -1, \pm2, \pm3$ are also shown for reference, but these do not appear in the spectral response in Fig.~\figref{fig:modes}{a}.

\section*{Floquet magnons of a magnetic vortex state}

To probe the dynamics far from equilibrium, a second microwave signal with $f_\mathrm{g}=\SI{200}{\mega\hertz}$ (at \SI{-5}{dBm}), close to the fundamental gyration frequency for the \SI{2}{\micro\meter}-wide disk, is applied in addition to the first microwave signal with $f_{nm}=\SI{6.2}{\giga\hertz}$ that excites the $(0,1)$ mode. In both the experimental and simulated spectra, a frequency comb appears around the initially excited azimuthal mode ($f_{nm}$), with the spacing between the sideband peaks given by the gyration frequency $f_\mathrm{g}$. In the simulated spectra, the gyration along with its harmonics are visible in the sub-GHz range, but these are below the instrumental limit of our experiment. Fig.~\figref{fig:modes}{d} shows the simulated spatial profiles of the modes that constitute the frequency comb.  Neighboring modes in the comb are not only shifted in frequency by $\pm f_\mathrm{g}$, but their azimuthal index also vary by an increment of $\Delta m = \pm 1$. Importantly, the magnon modes about the gyrating vortex exhibit qualitatively different profiles compared to their counterparts in the static case [Fig.~\figref{fig:modes}{b}], indicating a fundamental change in character resulting from the periodic vortex motion.

Previous observations of magnon frequency combs have been attributed to resonant three- or four-magnon scattering involving regular modes of the system~\cite{hula_spin-wave_2022}, including off-resonant scattering within the linewidths of the existing modes\cite{wang_twisted_2022}, or scattering with other textures such as skyrmions~\cite{wang_magnonic_2021} or domain walls~\cite{zhou_spin_2021}. The modes we observe within the frequency comb are not part of the regular magnon spectrum. This is confirmed in micromagnetic simulations of the Langevin dynamics of the magnetization in which thermal fields populate the magnon modes. In the absence of microwave drive, we recover the regular spectrum of vortex eigenmodes corresponding to a static core, as shown in Fig.~\figref{fig:modes}{e} for the four lowest radial indices, $n=0$ to $3$. Higher-order azimuthal modes in this configuration are typically degenerate, while for small $m$ ($\pm 1, \pm 2$ for the  disk dimensions studied) the magnon modes are hybridized with the gyrotropic mode and exhibit a sizeable frequency difference between opposite azimuthal numbers~\cite{guslienko_dynamic_2008}. We can compare this regular magnon dispersion to the results obtained for a gyrating vortex, by overlaying the modes identified in the frequency comb as hollow blue dots in Fig.~\figref{fig:modes}{e}. It is clear that several modes in the frequency comb do not coincide with the regular magnon dispersion.

The frequency comb appears when the disk is excited with two frequencies $f_{nm}$ and $f_\mathrm{g}$ simultaneously. The additional low-amplitude, low-frequency drive at $f_\mathrm{g}$ leads to a periodic modulation of the ground state which results in the generation of Floquet states. This behavior is reproduced in micromagnetics simulations when a rotating in-plane magnetic field, whose frequency matches the gyrotropic mode frequency $f_\mathrm{g}$, is included in addition to the thermal fluctuations. As Fig.~\figref{fig:modes}{f} shows, the resulting spectra exhibit a frequency comb related to the Floquet magnon bands induced by the core gyration. For larger values of $\abs{m}$, these Floquet bands resemble the regular magnon dispersion but shifted by $\pm f_\mathrm{g}$ and $\pm m=1$. For smaller values of $\abs{m}$, however, the bands are much more complex, differing strongly from the regular magnon dispersion with band crossings and avoided level crossings. Furthermore, Floquet magnons of opposite azimuthal mode indices at large $\abs{m}$ exhibit a larger frequency difference compared to regular modes.

\section*{Floquet theory}

\begin{figure}
    \includegraphics{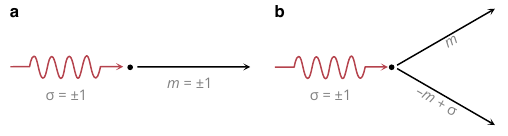}
    \caption{\textbf{Particle scattering representation.} Diagram of the \textbf{a}, two-particle and \textbf{b}, three-particle scattering between the gyrotropic mode of the vortex core gyration ($\sigma = \pm1$) and magnon modes with azimuthal indices $m$ and $m'$.}
    \label{fig:scattering}
    \end{figure}

The qualitative change to the magnon spectrum and the emergence of additional bands can be understood within Floquet theory of a many-particle picture of vortex modes that incorporates magnon-magnon interactions. Consider the following Hamiltonian describing vortex gyration and quantized magnon modes (with $\hbar = 1$), 
\begin{equation}\label{modelham}
    \hat{H} = \Omega \hat{N}_\sigma + \sum\limits_{nm} \omega_{nm} \hat{n}_{nm} + \hat{H}_\mathrm{int},
\end{equation}
with $\Omega = 2\pi f_\mathrm{g}$ and $\hat{N}_\sigma=\hat{A}^\dagger_\sigma\hat{A}_\sigma$ being the frequency and occupation number of the gyrotropic mode, $\sigma=\pm 1$ the gyration sense, and $\omega_{nm}$ and $\hat{n}_{nm}=\hat{a}^\dagger_{nm}\hat{a}_{nm}$ the respective quantities of the regular magnon modes. The operators $\hat{A}_\sigma^\dagger$ ($\hat{A}_\sigma$) and $\hat{a}_i^\dagger$ ($\hat{a}_i$) denote the bosonic creation (annihilation) operators of the modes and $\hat{H}_\mathrm{int}$ the interaction between the vortex gyration and the magnon modes. Moving into the Dirac picture with respect to the gyromode $\hat{A}_\sigma$ by transforming $\hat{A}\rightarrow \exp(-i\Omega t)\hat{A}$ allows to drop the term $\Omega \hat{N}$. In this picture, the terms of the interaction Hamiltonian $\hat{H}_\mathrm{int} = \hat{H}^\mathrm{(1)} + \hat{H}^\mathrm{(2)} + ...$, which are lowest in order of regular magnon modes, are given as 
%
\begin{align}
    \hat{H}^{(1)} & = e^{i\Omega t}\sum\limits_n U_{n\sigma}\hat{A}^\dagger_\sigma \hat{a}_{n\sigma} +  \mathrm{h.c.}\\
    \hat{H}^{(2)} & = e^{i\Omega t}\sum\limits_{nn^\prime m}    V_{nn^\prime m\sigma}  \hat{A}^\dagger_\sigma \hat{a}_{n'-m+\sigma}\hat{a}_{nm} + \mathrm{h.c.}
\end{align}
and describe two-particle and three-particle scattering as well as their time-reversed processes (given by the symbol "h.c." denoting the Hermitian conjugate of the preceding term), as sketched in Fig.~\ref{fig:scattering}. The parameters $U_{n\sigma}$ and $V_{nn^\prime \sigma}$ describe the coupling of the regular magnon modes $\hat{a}_i$ to a gyrating vortex core (represented by the gyration mode $\hat{A}_\sigma$), which is conceptually similar to the magnon scattering on a traveling magnetic domain wall. These parameters can be found qualitatively in a Lagrangian collective variables approach assuming a constant gyration radius.

Under steady-state gyration of the vortex core, the system described by $\hat{H}(t)$ becomes periodic in time, which allows us to apply the Floquet theorem. This states that the spectrum in a time-periodic system can be obtained from the Floquet Hamiltonian $\hat{H}_\mathrm{F}$, which enters in the total time-evolution operator,
\begin{align}
    \hat{U}(t_2,t_1) = e^{-\textrm{i}\hat{K}(t_2)}e^{-\textrm{i}\hat{H}_F(t_1-t_2)}e^{\textrm{i}\hat{K}(t_1)}.
\end{align}
Here, $\hat{K}(t) = \hat{K}(t+T_\mathrm{g})$ is the kick operator which is time-periodic with period $T_\mathrm{g}=2\pi/\Omega$, see, e.g., Refs.~ \citenum{goldman_periodically_2014,bukov_universal_2015,Eckardt_high-frequency_2015}. Consequently, the Floquet spectrum of the system is only defined up to a multiple of the gyration frequency $\Omega$ and, therefore, can be indexed with an additional mode index $\lambda$. 
For the present model $\hat{K}(t)$ and $\hat{H}_\mathrm{F}$ can be found analytically, resulting in the Floquet spectrum 
\begin{equation}
    \omega_{nm\lambda} = \omega_{nm}^\prime + \omega_0 + \lambda\Omega \quad \text{with}\ \lambda = 0, \pm 1, \pm 2, ... 
\end{equation}
with $\omega_0$ being a constant frequency shift due to the slight change in the ground state energy with respect to the vortex with a static core. 
Importantly, the frequencies $\omega_{nm}^\prime \neq \omega_{nm}$ of the new modes do not coincide with the frequencies of the original magnon modes $\omega_{nm}$ but include non-perturbative corrections due to magnon-magnon interactions. This model explains the appearance of additional modes in the magnon spectrum of a gyrating vortex seen in Fig.~\figref{fig:modes}{f}. Moreover, magnon-magnon interactions with the gyration result in the strong deviations of the Floquet branches from being mere $\Omega$-shifted copies of the original dispersion. This becomes apparent by overlaying the Floquet spectrum over the regular ground-state magnon dispersion from Fig.~\figref{fig:modes}{e}, where large deviations can already be seen in the zeroth-order branches $\lambda = 0$. Such renormalizations are a characteristic signature of Floquet systems~\cite{dunlap_dynamical_1986, grifoni_driven_1998, goldman_periodically_2014, bukov_universal_2015, Eckardt_high-frequency_2015} and cannot be accounted for by simple frequency multiplication.  

Consider the coupling between the core gyration and magnon modes within the usual particle picture, as illustrated in Fig.~\ref{fig:scattering}. The gyration involves selection rules for the azimuthal mode index, $m$, by imposing a difference of $\pm1$ in the scattered mode indices. This is reminiscent of \textit{Umklapp} processes in crystals, where momentum conservation is satisfied up to the reciprocal space vector, \textbf{G}, i.e. $\textbf{k} = \textbf{k'} + \textbf{G}$, with the initial and scattered wave vectors $\textbf{k'}$ and $\textbf{k}$,  respectively. Here, we observe something analogous for the azimuthal mode number, $m = m' + \sigma$, which indicates that the gyration plays the role of a reciprocal space vector, albeit limited to values of $\pm1$, and shifts the Floquet bands not only in energy but also in the azimuthal mode index.

\section*{Transient dynamics of Floquet magnons}

    \begin{figure*}
    \includegraphics[width=16cm]{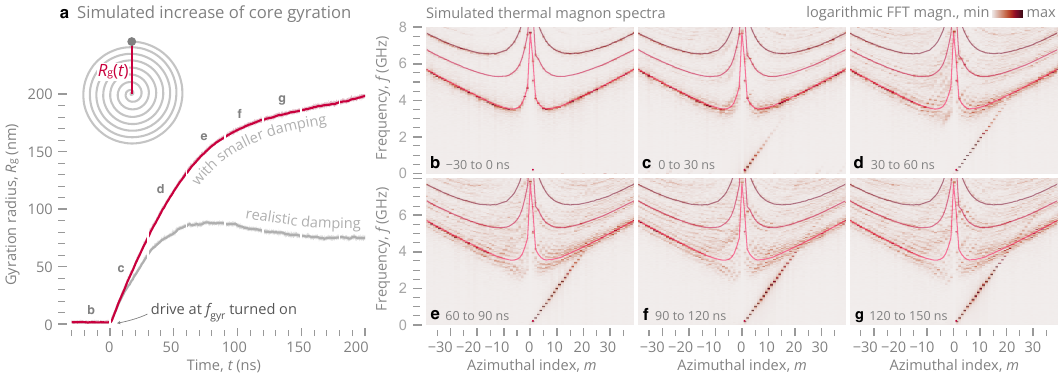}
    \caption{\textbf{Emergence of Floquet bands with increasing core gyration radius.}  \textbf{a}, Simulation data for a \SI{2}{\micro\meter}-wide disk showing the temporal evolution of the gyration radius $R_\mathrm{g}$ when driven with its resonance frequency $f_\mathrm{g}$ at room temperature, for two different values of the magnetic damping. A running average is shown and placed on top of the raw data. \textbf{b}-\textbf{g}, Micromagnetic simulations show the emergence of Floquet bands in the thermal magnon spectrum of a gyrating vortex. The evolution of the thermal magnon spectra is plotted for the different time segments as annotated in \textbf{a}. To better discriminate the different dispersion branches, the magnetic damping has been artificially reduced to $\alpha=0.0001$ which leads to a longer transient of the forced core gyration, compared to a more realistic value of $\alpha=0.007$.}
    \label{fig:transient}
    \end{figure*}

Like most studies on Floquet engineering to date, the theoretical framework above describes how new magnon bands are generated under periodic drive in the steady-state, i.e., for a constant gyration radius $R_\mathrm{g}$. However, the transient dynamics related to these bands remains largely unexplored. Figure~\ref{fig:transient} illustrates how these bands emerge from an initial static state in a \SI{2}{\micro\meter}-wide Ni$_{81}$Fe$_{19}$ disk using micromagnetics simulations. We used a small damping constant of $\alpha=0.0001$ (compared to the more realistic value of $\alpha=0.007$) in order to obtain narrower spectral lines and longer transients toward steady-state gyration. After an initial thermalization step in which the Langevin dynamics is simulated over \SI{230}{\nano\second}, a rotating in-plane magnetic field at frequency $f_\mathrm{g}=\SI{200}{\mega\hertz}$ is applied to excite the vortex gyration. The core gyration radius as a function of time is shown in Fig.~\figref{fig:transient}{a}, while the magnon spectra obtained over several successive intervals are shown in Figs.~\figref{fig:transient}{b-g}.

Before the onset of gyration ($t<\SI{0}{\nano\second}$, Fig.~\figref{fig:transient}{b}), the vortex core undergoes low-amplitude Brownian motion close to the disk center, with a thermal magnon spectrum corresponding to the case shown in Fig~\figref{fig:modes}{e}. As the rotating field is switched on and the gyration radius $R_\mathrm{g}$ increases, the Floquet bands emerge progressively as witnessed in the different snapshots of the dispersion relations in Fig.~\figref{fig:transient}{b} to \figref{fig:transient}{g}. The gradual appearance of the Floquet bands is correlated with the growth in the gyration orbit, which further underscores the key role of core gyration, rather than the external driving field.

\section*{Self-induced Floquet magnons }

    \begin{figure*}
    \includegraphics[width=16cm]{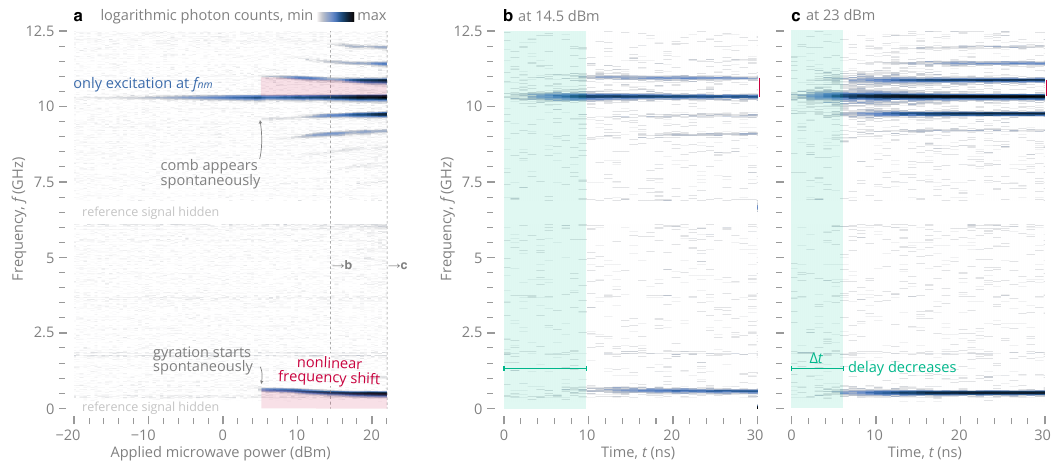}
    \caption{\textbf{Self-induced Floquet magnon bands at high driving powers.} \textbf{a}, Micro-focused BLS spectra showing the emergence of Floquet magnon bands with increasing applied microwave power when only a single azimuthal mode ($0,1$) is excited at $f_{nm}=\SI{10.2}{\giga\hertz}$.  \textbf{b},\textbf{c} Time-resolved BLS microscopy measurements demonstrate the spontaneous onset of the Floquet bands. The higher the excitation power of the azimuthal mode, the faster the vortex gyration sets in and Floquet states emerge. To be able to measure the frequency of the core gyration directly, the measurements shown here were performed on a Ni$_{81}$Fe$_{19}$ disk with \SI{500}{\nano\meter} diameter and \SI{50}{\nano\meter} thickness in which the gyration frequency is larger and  allows its detection with micro-focused BLS at the same time as the frequency comb.}
    \label{fig:spontaneous}
    \end{figure*}

We found that these Floquet states can also be self-induced by exciting magnon modes directly, using microwave frequencies an order of magnitude above the gyration frequency. We demonstrate this experimentally in a \SI{500}{\nano\meter}-diameter, \SI{50}{\nano\meter}-thick disk, for which the fundamental gyration frequency is around $f_\mathrm{g}= \SI{500}{\mega\hertz}$. This value is above the instrumental limit of our micro-focus BLS measurements, which allows us to probe the gyration and magnon modes simultaneously. Figure~\figref{fig:spontaneous}{a} shows the measured BLS spectra as a function of the microwave power when a single frequency of $f_{nm}=\SI{10.2}{\giga\hertz}$ is applied. At low power, the microwave field resonantly excites a single azimuthal mode, as shown in Fig.~\figref{fig:spontaneous}{a}. At the threshold power of about \SI{5}{\decibel}m, a frequency comb around this azimuthal mode appears, along with a strong spectral response in the sub-GHz regime associated with core gyration. Above this threshold, increases in the microwave power lead to a reduction in the gyration frequency, a phenomenon known as nonlinear redshift~\cite{Guslienko_PhysRevB.82.014402, sushruthElectricalMeasurementMagneticfieldimpeded2016}. This redshift is imprinted in the frequency spacing of the Floquet states [red shaded areas in Fig.~\figref{fig:spontaneous}{a}]. Additionally, increasing powers also result in shorter delays ($\Delta t$) before the onset of gyration, as observed in the time resolved BLS spectra in Figs.~\figref{fig:spontaneous}{b,c}. This power dependence is a hallmark of nonlinear mode coupling, which in the present case involves the parametric excitation of the gyrotropic mode by an azimuthal magnon mode.

While the connection between frequency combs and core gyration has been discussed in another context~\cite{wang_twisted_2022}, our experiments establish a clear connection to Floquet physics that has previously been overlooked. The Floquet mechanism here also differs crucially from previous studies in that the primary source of modulation (core gyration) is not excited directly, but rather through the nonlinear coupling to other modes which are populated by the external drive. This self-induced mechanism is a novel feature of the vortex state.

\section*{Conclusion and outlook}

We have demonstrated a new approach to Floquet engineering in which the periodic drive of an internal mode, namely vortex core gyration, induces Floquet magnon bands through the nonlinear coupling with this mode. This differs inherently from more traditional approaches in which laser or microwave illumination is used to periodically modulate material constants such as exchange or anisotropy. Another difference lies in the fact that the Floquet bands can be induced either by driving the core gyration directly, with an oscillatory external magnetic field in the sub-GHz range, or indirectly, through the excitation of a higher-order eigenmode in the GHz range, which couples to the gyration through nonlinear interactions. This suggests new avenues to explore Floquet engineering with other topological magnetic solitons, such as domain walls and skyrmions, which also possess low-frequency Goldstone-like modes that couple to high-frequency eigenmodes. We anticipate that this paradigm may find applications beyond magnetism, such as in ferroelectric or superconducting systems, which also host solitonic objects like vortices.

\bibliographystyle{Science}
\bibliography{references.bib}


\section*{Acknowledgments}
The authors thank A. Manchon for pointing us in the direction of Floquet physics, B. Scheumann for depositing the Ni$_{81}$Fe$_{19}$ and Au films, V. Iurchuk for contributing to the micromagnetic simulations in the early phase of the project, and A. Hoffmann for fruitful discussions.

\subsection*{Funding}
This work was supported by the Deutsche Forschungsgemeinschaft (DFG) through the programs KA 5069/3-1 and GL 1041/1-1, and the EU Research and Innovation Programme Horizon Europe under grant agreement no. 101070290 (NIMFEIA). Support by the Nanofabrication Facilities Rossendorf (NanoFaRo) at the IBC is gratefully acknowledged. 

\subsection*{Author contributions}

H.S., K.S. and C.H. conceived the experiments. K.S. fabricated the sample. C.H. carried out the experiments. C.H. and J.-V.K. performed the micromagnetic simulations. J.M., L.K. and J.-V.K. developed the theory. L.K., C.H., J.-V.K, K.S. and H.S. visualized the results. J.-V.K., T.D., J.M., A.K., K.S., J.F., and H.S. acquired funding. All authors analyzed the data and discussed the results. L.K., J.-V.K., J.M., A.K., and K.S. wrote the original draft of the paper. All authors reviewed and edited the paper.

\subsection*{Competing interest}
The authors have no conflicts to disclose.

\subsection*{Data and materials availability}
The data that support the findings of this study are openly available in RODARE. We used scientific color maps to prevent visual distortion of the data and exclusion of readers with color-vision deficiencies~\cite{crameri_misuse_2020}. Specifically, the color maps used include \textit{oslo} (https://www.fabiocrameri.ch/colourmaps/), \textit{cmocean.map} (https://matplotlib.org/cmocean/) and \textit{guppy} (https://cmasher.readthedocs.io/index.html~\cite{van_der_velden_cmasher_2020}).

\section*{Supplementary materials}
Materials and Methods

Figure S1

References [51-58]

\clearpage
\newpage
\renewcommand{\theequation}{S\arabic{equation}}
\renewcommand{\thefigure}{S\arabic{figure}}
\setcounter{equation}{0}
\setcounter{figure}{0}

\section*{Supplementary materials} 

\subsection*{Sample preparation}

\begin{figure}[b]
     \includegraphics{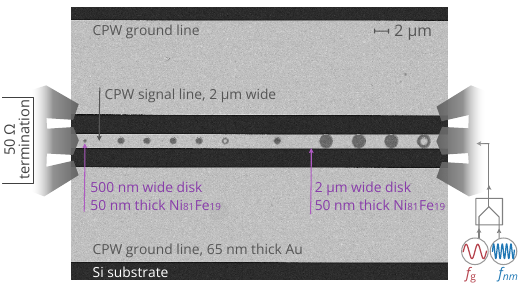}
     \caption{\textbf{Scanning electron micrograph of the studied sample.}   $\SI{50}{\nano\meter}$-thick Ni$_{81}$Fe$_{19}$ disks with different diameters $D=0.5, 1$, and $\SI{2}{\micro\meter}$ are positioned on top of the central signal line of a coplanar waveguide (CPW). For actively exciting magnetization dynamics in different frequency ranges $f_g$ and $f_{nm}$, two separate signal sources are used.}
     \label{fig:sample}
     \end{figure}

The experiments discussed in this work were performed on the sample shown in Fig.~\ref{fig:sample} that was fabricated in a two-step procedure. We started with an undoped, high-resistance silicon substrate. In a first step, we patterned the coplanar wave\-guide using a double-layer resist of methyl methacrylate (EL11) and poly(methyl methacrylate) (950 PMMA-A2), electron beam lithography, electron beam evaporation of a Cr($\SI{5}{\nano\meter}$)/Au($\SI{65}{\nano\meter}$) layer and subsequent lift-off in an acetone bath. The central line and ground lines of the coplanar waveguide have widths of $\SI{2}{\micro\meter}$ and $\SI{13.5}{\micro\meter}$, respectively, the gap between them is $\SI{2.8}{\micro\meter}$ wide.

In a second step, the magnetic structures with different diameters ($\SI{500}{\nano\meter}$, $\SI{1}{\micro\meter}$, and $\SI{2}{\micro\meter}$) are patterned directly on top of the central signal line of the coplanar waveguide. Therefore, we use a poly(methyl methacrylate) (950 PMMA-A6) resist, electron beam lithography, electron beam evaporation of a Cr($\SI{5}{\nano\meter}$)/Ni$_{81}$Fe$_{19}(\SI{50}{\nano\meter}$)/Cr($\SI{2}{\nano\meter}$) layer and subsequent lift-off in an acetone bath.

\subsection*{Time-resolved Brillouin light scattering microscopy} 

All experimental measurements were performed at room temperature. The magnon spectra were detected by means of Brillouin light scattering microscopy~\cite{sebastian_micro-focused_2015}. Therefore, a monochromatic, continuous-wave $\SI{532}{\nano\meter}$ laser was focused onto the sample surface using a microscope lens with a high numerical aperture, yielding a spatial resolution of about $\SI{300}{\nano\meter}$. The backscattered light was then directed into a Tandem Fabry-P\'{e}rot interferometer~\cite{mock_construction_1987} in order to measure the frequency shift caused by the inelastic scattering of photons and magnons. The detected intensity of the frequency-shifted signal is directly proportional to the magnon intensity at the respective focusing position. 

The low-frequency range is challenging to measure using Brillouin light scattering due to the high intensity of the elastically scattered Rayleigh peak, whose flank covers the low frequency signals. In details, the low-frequency detection limit of the interferometer is determined by the course spacing of the etalon mirrors (distance $L_1$ in Fig.~1(b) in Ref.~\citenum{mock_construction_1987}) which defines the free spectral range. The data shown in Fig.~\ref{fig:spontaneous} was recorded with a mirror spacing of \SI{23}{\milli\meter}, 
while the spectra plotted in \ref{fig:modes} were measured with a mirror spacing of \SI{18}{\milli\meter}. 
Hence, the different detection limits in the low-frequency range.

When using microwave pulses to excite magnons, it is possible to measure the temporal evolution of the magnon spectra using a time-of-flight principle. Therefore, we simultaneously monitor the state of the interferometer and the time when each photon is detected with respect to a clock provided by the microwave generator using a time-to-digital converter. In order to acquire enough signal, the pulsed experiment needs to be repeated stroboscopically, covering hundred thousands of repetitions. 

During all experiments, the investigated  microstructure was imaged using a red LED and a CCD camera. Displacements and drifts of the sample were tracked by an image recognition algorithm and compensated by the sample positioning system.

To account for the different spatial distributions of the magnon modes, the signal was integrated over 3 radial and 4 azimuthal positions across half the \SI{2}{\micro\meter}-wide disk. The signal for the \SI{500}{\nano\meter}-wide disk was obtained at a singular position.

\subsection*{Micromagnetic simulations}

Simulations of the vortex dynamics were performed using the open-source finite-difference micromagnetics code \textsc{MuMax3}~\cite{vansteenkiste_design_2014}, which performs a time integration of the Landau-Lifshitz-Gilbert equation of motion of the magnetization $\bm{m}(\bm{r},t)$,
\begin{equation}
    \pdv{\bm{m}}{t} = -\gamma \bm{m} \times \qty(\bm{B}_\mathrm{eff} + \bm{b}_\mathrm{th}) + \alpha \bm{m} \times \pdv{\bm{m}}{t}.
\end{equation}
Here, $\bm{m}(\bm{r},t) = \bm{M}(\bm{r},t)/M_s$ is a unit vector representing the orientation of the magnetization field $\bm{M}(\bm{r},t)$ with $M_s$ being the saturation magnetization, $\gamma = g\mu_B/\hbar$ is the gyromagnetic constant, and $\alpha$ is the dimensionless Gilbert-damping constant. The effective field, $\bm{B}_\mathrm{eff} = -\delta U/\delta \bm{M}$, represents a variational derivative of the total magnetic energy $U$ with respect to the magnetization, where $U$ contains contributions from the Zeeman, nearest-neighbor Heisenberg exchange, and dipole-dipole interactions. The term $\bm{b}_\mathrm{th}$ represents a stochastic field with zero mean, $\langle b_\mathrm{th}^i(\bm{r},t) \rangle = 0$ and spectral properties satisfying~\cite{brown_thermal_1963}
\begin{equation}
\langle b_\mathrm{th}^i(\bm{r},t) b_\mathrm{th}^j(\bm{r}',t')  \rangle = \frac{2\alpha k_B T}{\gamma M_s V} \delta_{ij} \delta(\bm{r}-\bm{r}') \delta(t-t') ,
\end{equation}
with amplitudes drawn from a Gaussian distribution. Here, $k_B$ is Boltzmann's constant, $T$ is the temperature, and $V$ denotes the volume of the finite difference cell. This stochastic term models the effect of thermal fluctuations acting on the magnetization dynamics. An adaptive time-step algorithm based on a sixth-order Runge-Kutta-Fehlberg method was used to perform the time integration~\cite{leliaert_adaptively_2017}.

We model our 50-nm thick, 2-$\mu$m diameter disk using $512 \times 512 \times 8$ finite difference cells with $\gamma=\SI{1.86e-11}{\radian/(Ts)}$, $M_\mathrm{s}=\SI{775}{\kilo\ampere/\meter}$, an exchange constant of $A_\mathrm{ex}=\SI{12}{\pico\joule/\meter}$, and $\alpha = 0.007$ –– the nominal value for this material. Note that smaller values of $\alpha$ were used to highlight different aspects of the Floquet bands, as discussed in the main text. 

The dispersion relations shown in Fig.~\ref{fig:modes} were computed as follows. For each of the non-driven and driven cases, time integration of the stochastic dynamics with $\alpha=0.0007$ was performed over an interval of $\SI{10}{\micro\second}$ and the out-of-plane magnetization fluctuations corresponding to the different azimuthal index, $a_m(t)$, were recorded. This involved a spatial Fourier decomposition that is computed on-the-fly by projecting out the magnetization $m_z(\bm{r},t)$ using the basis functions $\psi(\bm{r}) = e^{i m \phi}$, 
\begin{equation}
    a_m(t) = \int dV \; \psi^{*}(\bm{r}) \, m_z(\bm{r},t),
\end{equation}
with $\phi$ representing the angular variable in cylindrical coordinates. The power spectrum for each $a_m(t)$ was then computed using the Welch method, which involves averaging over the power spectra generated from the discrete Fourier transform of half-overlapping 400-ns Hann windows into which the original time series data is sliced. Note that the basis functions chosen are taken to be uniform across the film thickness, which means the power spectra shown only capture symmetric thickness modes. In the driven case, a rotating in-plane magnetic field with a frequency of 200 MHz and an amplitude of 0.05 mT was applied.

The dispersion relations shown in Fig.~\ref{fig:transient} were obtained in a similar way with $\alpha = 0.0001$, except that smaller 30-ns windows were used for the Fourier transform of the time series data in order to produce different snapshots in time of the Floquet bands.


\subsection*{Floquet theory}

To describe self-induced Floquet magnon modes in a magnetic vortex, we consider the following Hamiltonian (Eq.~\eqref{modelham} of the main text) 
\begin{align}
\hat{H} = \Omega \hat{N}_\sigma + \sum_{nm}\omega_{nm} \hat{n}_{nm} + \hat{H}_\text{int}
\end{align}
where $\Omega_\sigma$ is the frequency of the vortex-core oscillation with mode occupation $\hat{N}_\sigma=\hat{A}^\dagger_\sigma\hat{A}_\sigma$ and gyration sense (circular polarization) $\sigma=\pm1$. $\omega_{nm}$ is the regular magnon dispersion with mode occupation $\hat{n}_{nm}=\hat{a}^\dagger_{nm}\hat{a}_{nm}$, with $\hat{a}_{nm}$ ($\hat{a}^\dagger_{nm}$) a annihilation (creation) operator for magnons with radial index $n$ and azimuthal index $m$. $\hat{H}_\text{int}$ describes the coupling between the magnon modes and the vortex gyration. Phenomenologically, this interaction can be derived from the Lagrangian formulation in collective coordinates~\cite{bouzidi_motion_1990}, while here, for simplicity, we only consider a minimal model consistent with a Hamiltonian formulation of nonlinear magnon modes~\cite{krivosik_hamiltonian_2010, verba_theory_2021}. In leading order, this model is determined by terms only involving one and two magnon modes: $\hat{H}_\text{int} = \hat{H}^{(1)} + \hat{H}^{(2)}$, with
\begin{align}
\hat{H}^{(1)} & = \sum_{n} U_{n\sigma} \hat{A}^\dagger_\sigma\hat{a}_{n\sigma} + h.c. \\
\hat{H}^{(2)} &= \sum_{nn'm} V_{nn'm\sigma} \hat{A}^\dagger_\sigma \hat{a}_{n'-m+\sigma}\hat{a}_{nm} + h.c.
\end{align}
Note that we do not include a summation over $\sigma$ since, in practice, no superposition of left and right rotating vortex cores arise. $U_{n\sigma}$ and $V_{nn’m\sigma}$ are the expansion coefficients for the terms linear and quadratic in magnon operators.  We are interested in the self-induced changes of the spectrum $\omega_{nm}$ of the magnon modes due to the periodic oscillation of the core gyration. To this end it is convenient to go into the interaction picture with respect to the $H_0=\Omega \hat{N}_\sigma$. This results in replacing $\hat{A}^\dagger \rightarrow \hat{A}^\dagger \exp(\mathrm{i}\Omega t)$, $\hat{A} \rightarrow \hat{A} \exp(\mathrm{-i}\Omega t)$, while $H_0$ drops out. Then we are left with solving: 
\begin{align}
\hat{H}_{m} & = \hat{H}_m^{(0)}+\hat{H}_m^{(1)}+\hat{H}_m^{(2)}\\
\hat{H}_m^{(0)} & = \omega_m \hat{n}_{m}  \\
\hat{H}_m^{(1)} & = \delta_{m\sigma}\left(U_{\sigma}e^{\mathrm{i}\Omega t}  \hat{A}^\dagger_\sigma\hat{a}_{\sigma} + U^{*}_{\sigma}e^{-\mathrm{i}\Omega t} \hat{A}_\sigma\hat{a}^\dagger_{\sigma}\right) \\
\hat{H}_m^{(2)} & = V_{m}e^{\mathrm{i}\Omega t}   \hat{A}^\dagger_\sigma\hat{a}_{-m+\sigma}\hat{a}_{m} + V^*_{m}e^{-\mathrm{i}\Omega t}  \hat{A}_\sigma\hat{a}^\dagger_{-m+\sigma}\hat{a}^\dagger_{m},
\end{align}
where we dropped the indices $n$ to simplify the notation. Since this Hamiltonian is periodic in time, we can use the Floquet theorem, which states that the spectrum in the presence of driving can be obtained from the Floquet Hamiltonian $\hat{H}_F$ that enters the total evolution operator $\hat{U}(t_2,t_1)$ of the system as 
\begin{align}
\hat{U}(t_2,t_1) = e^{-\textrm{i}\hat{K}(t_2)}e^{-\textrm{i}\hat{H}_F(t_1-t_2)}e^{\textrm{i}\hat{K}(t_1)},
\end{align}
where $\hat{K}(t)=\hat{K}(t+T)$, is an hermitian operator that is periodic with period $T=2\pi/\Omega$. Clearly, the Floquet spectrum of the system is then defined only modulo the driving frequency $\Omega$. Interestingly, for the problem at hand we can find $\hat{H}_F$ and $\hat{K}$ analytically. Choosing $\hat{K}(t)=-\Omega \hat{n}_{m} t$ yields the time-independent Floquet Hamiltonian:
\begin{align}
\hat{H}_{F}^{(0)} & = \frac{1}{2} (\omega_m-\Omega)\hat{n}_{m} + \frac{1}{2} (\omega_{-m+\sigma}-\Omega)\hat{n}_{-m+\sigma}\\
\hat{H}_{F}^{(1)} & =  \frac{1}{2} (\delta_{m,\sigma}+\delta_{-m+\sigma,\sigma})\left(U_{\sigma} \hat{A}^\dagger_\sigma\hat{a}_{\sigma} + U^{*}_{\sigma}\hat{A}_\sigma\hat{a}^\dagger_{\sigma}\right) \\
\hat{H}_{F}^{(2)} & =  V_{m}\hat{A}^\dagger_\sigma\hat{a}_{-m+\sigma}\hat{a}_{m} + V^*_{m}\hat{A}_\sigma \hat{a}^\dagger_{-m+\sigma}\hat{a}^\dagger_{m}
\end{align}
For stroboscopic times $t_2=t_1+kT$, $k$ an integer, this is equivalent to the interaction picture with the Hamiltonian $\hat{H}'=\Omega\hat{n}_m$. To gain qualitative insight, it is sufficient to limit the remaining discussion to the case in which we can treat the steady core gyration as a classical variable, $A=\langle\hat{A}\rangle$, which we absorb in the definitions of the coupling strengths $U\rightarrow AU$ and $V\rightarrow AV$. The Floquet Hamiltonian can then be diagonalized with a generalized Bogolyubov transformation:
\begin{align}
\hat{\alpha}_m & = u_m(\hat{a}_m+\lambda_m) + v_m(\hat{a}^\dagger_{-m+\sigma}+\mu_m^*), \\
\hat{\alpha}_{-m+\sigma} & = u_m(\hat{a}_{-m+\sigma}+\mu_m) + v_m(\hat{a}^\dagger_m+\lambda_m^*),
\end{align}
where $\lambda_m,\mu_m$ and $u_m,v_m$, with $|u_m|^2-|v_m|^2=1$,  are (generally complex) parameters. Substitution and rearrangement of the terms under the sum yields the Floquet Hamiltonian
\begin{widetext}
\begin{align}
\hat{H}_{Fm} &= \omega^\prime_m\hat{\alpha}^\dagger_m\hat{\alpha}_m + \omega_m^0,\\
\omega^\prime_m &= \frac{1}{2}\left(\omega_m - \omega_{-m+\sigma}\right) + \frac{1}{2}\sqrt{(\omega_m + \omega_{-m+\sigma} - 2\Omega)^2-16|V_m|^2}\\
\omega^0_m &= -\frac{\frac{1}{2}(\omega_m - \Omega)|\delta_{-m+\sigma,\sigma}U_\sigma|^2 + \frac{1}{2}(\omega_{-m+\sigma}-\Omega)|\delta_{m,\sigma}U_\sigma|^2 }{\frac{1}{4}(\omega_m-\Omega)(\omega_{-m+\sigma}-\Omega)-|V_m|^2} -\frac{1}{2}\sqrt{(\omega_m + \omega_{-m+\sigma}-2\Omega)^2-16|V_m|^2}
\end{align}
\end{widetext}
Reinserting the index $n$, the Floquet spectrum of the system features the modes 
\begin{align}
\omega_{nm\lambda} &= \omega_{nm}^\prime + \omega_{nm}^0 + \lambda\Omega_{\sigma}
\end{align}
with $\lambda=0,\pm1,\pm2, \ldots$.

\end{document}